\documentclass[aps,prb,reprint,
			   groupedaddress,superscriptaddress,notitlepage,
			   amsfonts,amssymb,amsmath,
			   citeautoscript,
			   a4paper]{revtex4-2}

\usepackage[pdftex]{hyperref}
\hypersetup{colorlinks,
			linkcolor={blue!75!black!80!yellow},
			citecolor={blue!75!black!80!yellow},
			urlcolor={blue!75!black!80!yellow},
			pdfstartview=FitH}
\usepackage{graphicx}
\usepackage{mathrsfs}
\usepackage{amsthm}
\usepackage{xspace}
\usepackage{braket}
\usepackage{xr}
\usepackage{float} % % fix figure position
\usepackage{gensymb} % to use \degree command
\usepackage{xcolor,soul}
\usepackage{stmaryrd} % provides \llbracket and \rrbracket for difference-operations
\usepackage[UKenglish]{babel}
\usepackage{placeins} % provides \FloatBarrier
\usepackage{amsmath,amssymb}
\usepackage{siunitx}
\DeclareSIUnit[number-unit-product=]\percent{\char`\%} % remove space before percentage "units"

\usepackage{txfonts}  %Times Roman fonts
\usepackage{txfontsb} %Addition for txfonts, including old style numerals and greek

\newcommand{\iu}{\mathrm{i}}

\newcommand{\appropto}{\mathrel{\vcenter{
			\offinterlineskip\halign{\hfil$##$\cr
				\propto\cr\noalign{\kern2pt}\sim\cr\noalign{\kern-2pt}}}}}
\newcommand{\ie}{i.e.\@\xspace}  %Gobble-spaces of the "small" type ("small" via \@)

\newcommand{\eg}{e.g.\@\xspace}

\makeatletter
\newcommand*{\addFileDependency}[1]{% argument=file name and extension
  \typeout{(#1)}
  \@addtofilelist{#1}
  \IfFileExists{#1}{}{\typeout{No file #1.}}
}
\makeatother
\newcommand*{\myexternaldocument}[1]{%
    \externaldocument{#1}%
    \addFileDependency{#1.tex}%
    \addFileDependency{#1.aux}%
}
\myexternaldocument{SM}
\newcommand{\sisec}[1]{Sec.~\ref{#1}}
        
\newcommand{\mitaffil}{\footnotesize Research Laboratory of Electronics and Department of Physics, Massachusetts Institute of Technology, Cambridge, Massachusetts 02139, USA}
\newcommand{\ucbaffil}{\footnotesize Department of Physics, University of California, Berkeley, CA 94720, USA}

\begin{document}

\title{Three-dimensional non-Abelian generalizations of the Hofstadter model: spin-orbit-coupled butterfly trios}
\author{Vincent Liu}
\email{vincent\_liu@berkeley.edu}
\affiliation{\mitaffil}
\affiliation{\ucbaffil}
\author{Yi Yang}
\email{yiy@mit.edu}
\affiliation{\mitaffil}
\author{John D. Joannopoulos}
\affiliation{\mitaffil}
\author{Marin~Solja\v{c}i\'{c}}
\affiliation{\mitaffil}
\date{\today}

\begin{abstract}
    \noindent
    We theoretically introduce and study a three-dimensional Hofstadter model with linearly varying non-Abelian gauge potentials along all three dimensions. The model can be interpreted as spin-orbit coupling among a trio of Hofstadter butterfly pairs since each Cartesian surface ($xy$, $yz$, or $zx$) of the model reduces to a two-dimensional non-Abelian Hofstadter problem. By evaluating the commutativity among arbitrary loop operators around all axes, we derive its genuine (necessary and sufficient) non-Abelian condition, namely, at least two out of the three hopping phases should be neither 0 nor $\pi$. Under different choices of gauge fields in either the Abelian or the non-Abelian regime, both weak and strong topological insulating phases are identified in the model.
\end{abstract}

\maketitle

The Hofstadter model~\cite{hofstadter1976energy} is fundamental to the study of the quantum Hall effect and topology in condensed matter physics. It describes non-interacting electrons hopping in a two-dimensional square lattice under a perpendicular U(1) magnetic field. In solid state systems, the magnetic fields required for realizing the Hofstadter model had been inaccessible experimentally until the introduction of Moir\'{e} superlattices~\cite{dean2013hofstadter,ponomarenko2013cloning,hunt2013massive}, which expand the size of unit cells and the threaded magnetic flux substantially.   

An alternative way to realize the Hofstadter model in real space is via synthetic gauge fields~\cite{aidelsburger2018artificial} in artificial, engineered systems. So far, a plethora of realizations have been achieved, including microwave scatterers~\cite{kuhl1998microwave}, cold atoms~\cite{aidelsburger2013realization,miyake2013realizing}, acoustics~\cite{ni2019observation}, photons~\cite{hafezi2013imaging}, and superconducting qubits~\cite{roushan2017spectroscopic,owens2018quarter}. 

In two dimensions, there has been considerable interest in studying non-Abelian generalizations of the Hofstadter model, which replace the Abelian U(1) gauge fields with non-Abelian choices. Categorized by the spatial arrangements of the gauge fields, there have been theoretical studies that feature constant~\cite{osterloh2005cold,goldman2009ultracold,burrello2010non,kosior2014simulation} and linearly varying non-Abelian gauge fields in one~\cite{goldman2010realistic} or two~\cite{yang2020non} spatial dimensions. Experimentally, real-space building blocks of non-Abelian SU(2) gauge fields were demonstrated~\cite{yang2019synthesis} in photonics via minimal-scheme, non-Abelian Aharonov--Bohm interference~\cite{wu1975concept,chen2019non,zygelman2021topological}. In addition, non-Abelian braiding of topological zero modes were proposed and realized with coupled waveguide arrays~\cite{iadecola2016non,noh2020braiding}. These advances indicate a possibility to realize non-Abelian Hofstadter models in photonic systems. 

In three dimensions (3D), Abelian generalizations of the Hofstadter model corresponding to tilted magnetic fields have been studied as many as three decades ago~\cite{hasegawa1990generalized,montambaux1990quantized,kunszt1991electron,kohmoto1992diophantine}. In particular, such a 3D problem with arbitrarily oriented three-dimensional flux states can be reduced to a one-dimensional hopping in a suitably chosen gauge~\cite{kunszt1991electron}. Moreover, the model was found to support the 3D quantum Hall effect with quantized Hall conductance under anisotropic conditions~\cite{koshino2001hofstadter}. Far fewer non-Abelian generalizations have been considered in 3D. Specifically, topological properties of a 3D Hofstadter-like problem with non-Abelian gauge potentials that vary linearly along a single direction (of the remaining two directions, one has a constant SU(2) gauge and the other a real hopping) has been studied~\cite{li2015time}, which is shown to also be reducible to an effective 1D problem.

In this work, we theoretically propose and study a 3D non-Abelian Hofstadter model on a cubic lattice, whose non-Abelian SU(2) gauge potentials are linearly varying along all three dimensions. A crucial feature of our model construction is that any arbitrary Cartesian surface (either $xy$, $yz$, or $zx$) of our 3D model reduces to a two-dimensional non-Abelian Hofstadter model in the symmetric gauge~\cite{yang2020non}. Meanwhile, adjacent layers are coupled with spatially varying hopping. Therefore, the whole system can be treated as the spin-orbit coupling among three Hofstadter butterflies (each encoded along a single dimension). By evaluating the commutativity between arbitrary real-space loop operators, we derive the necessary and sufficient condition for our model to be genuinely non-Abelian, namely, that at least two out of the three hopping phases are neither 0 nor $\pi$. Compared to the 3D Abelian Hofstadter model, the spin-orbit coupling in the 3D non-Abelian Hofstadter model opens up new band gaps. We further show that these gaps can be of either weak or strong 3D $\mathbb{Z}_2$ topological insulating phases under different choices of the gauge potentials.

As warmup, we begin by describing homogeneous magnetic fields in 3D cubic lattices labeled by $(m,n,l)$. First, homogeneous U(1) magnetic fields in three dimensions can be described by three-dimensional Hofstadter models. In a general form that is akin to our proposed model below, the 3D Abelian Hofstadter model~\cite{hasegawa1990generalized,montambaux1990quantized,kunszt1991electron,kohmoto1992diophantine,koshino2001hofstadter} can be expressed as
\begin{align} \label{eq:abelian_3d}
    \begin{split}
        H_1 = - \sum_{m,n,l} t_x& c^\dagger_{m+1,n,l} e^{\iu \left(l \pm n \right) \theta_{x}} c_{m,n,l}  \\
        + t_y & c^\dagger_{m, n+1, l} e^{\iu \left( m \pm l \right) \theta_{y}} c_{m,n,l}  \\
        + t_z & c^\dagger_{m, n, l+1} e^{\iu \left(n \pm m \right) \theta_{z}} c_{m,n,l} + \mathrm{H.c.,} 
    \end{split}
\end{align}
where $\theta_{x}$, $\theta_{y}$, and $\theta_{z}$ are hopping phases and $t_x$, $t_y$, and $t_z$ are hopping amplitudes along three directions. $H_1$ corresponds to a gauge potential $\mathbf{A}_1= \left((l\pm n)\theta_{x},(m \pm l)\theta_{y},(n \pm m)\theta_{z}\right)$, which varies linearly along all three dimensions. The associated magnetic flux can be evaluated from the loop operator along different Cartesian surfaces, which yields a magnetic field of $\mathbf{B}_1=\left(\theta_{y}\mp\theta_{x},\theta_{x}\mp\theta_{z},\theta_{z}\mp\theta_{y}\right)$. Evidently, the choice of $\pm$ in Eq.~\ref{eq:abelian_3d} corresponds to different homogeneous magnetic fields (see Supplementary Materials Fig.~\ref{sm_fig:compfig}b and d).

On the other hand, a homogeneous non-Abelian SU(2) gauge potential and its associated Hamiltonian on a cubic lattice are given by
\begin{align} \label{eq:goldman_3d}
    \begin{split}
    H_2 = - \sum_{m,n,l}  t_x & c^\dagger_{m+1,n,l} e^{\iu \theta_{x} \sigma_x} c_{m,n,l} \\
            + t_y & c^\dagger_{m, n+1, l} e^{\iu \theta_{y} \sigma_y} c_{m,n,l} \\
            + t_z & c^\dagger_{m, n, l+1} e^{\iu \theta_{z} \sigma_z} c_{m,n,l} + \mathrm{H.c.,} 
    \end{split}
\end{align}
which corresponds to a gauge potential $\mathbf{A}_2= \left( \theta_{x} \sigma_x, \theta_{y} \sigma_y, \theta_{z} \sigma_z \right)$, where $\sigma_{x,y,z}$ are Pauli matrices. Every layer of $H_2$ becomes the celebrated 2D homogeneous non-Abelian model as proposed in Refs.~\cite{goldman2009non,goldman2010realistic}. The 3D non-Abelian gauge potential $\mathbf{A}_2$ describes a spatially homogeneous SU(2) magnetic field $\mathbf{B}_2=  \left( 2 \theta_{y} \theta_{z} \sigma_x, 2 \theta_{z} \theta_{x} \sigma_y, 2 \theta_{x} \theta_{y} \sigma_z \right).$ The associated Bloch Hamiltonian of $H_2$ is given by
\begin{align} \label{eq:h2bloch}
    h_2(\mathbf{k}) = 2\sum_i\cos k_i\cos \theta_i \sigma_0-2\sum_i\sin k_i\sin \theta_i \sigma_i
\end{align}
This two-band Hamiltonian is time-reversal-symmetric and always gapless. It hosts Weyl points at the eight time-reversal-invariant momenta (TRIMs).

\begin{figure}[htpb]
    \includegraphics[width=\columnwidth]{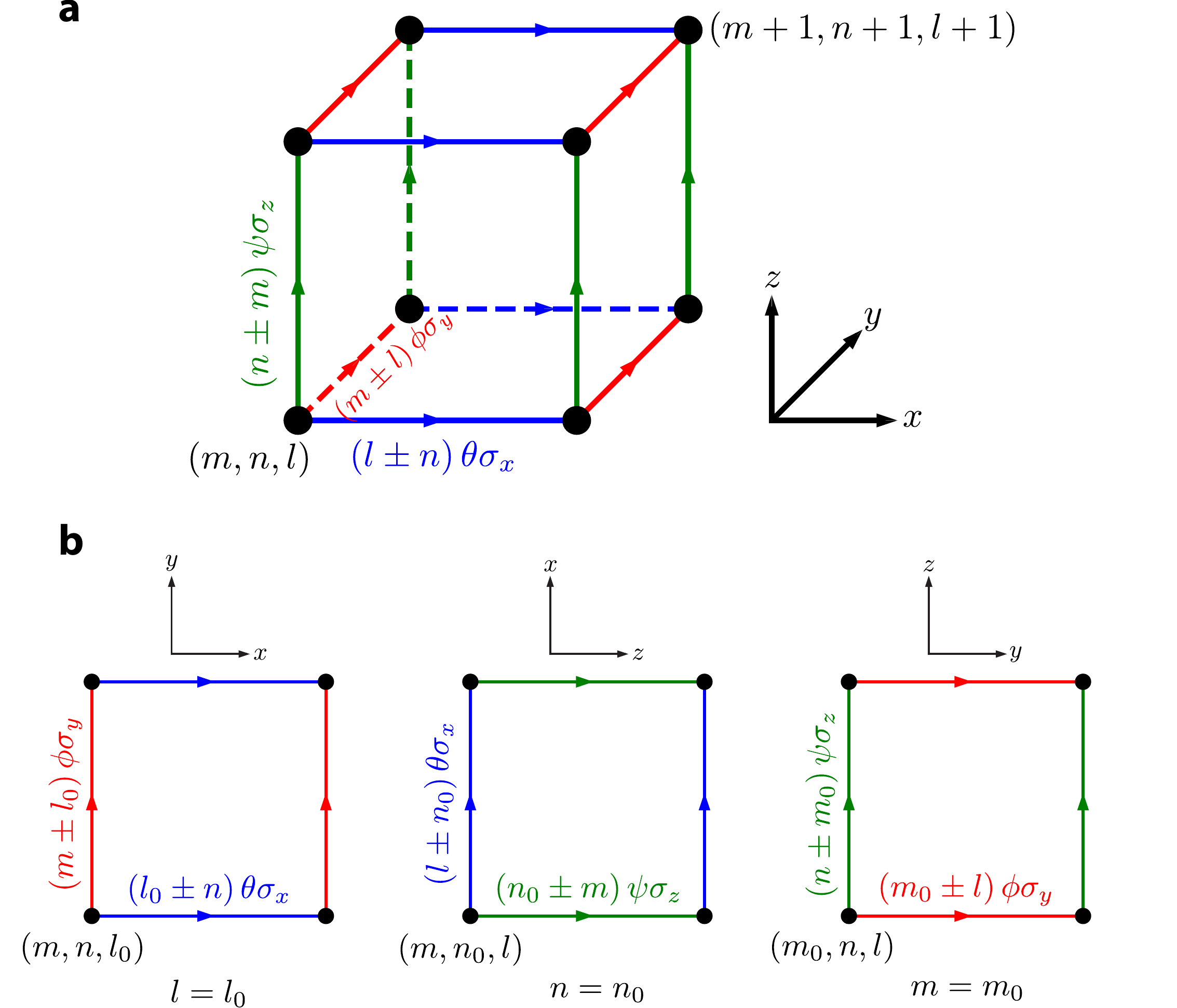}
    \caption{Three-dimensional non-Abelian generalization to the Hofstadter model: the blue, red, and green colors represent the link variables in the $x$, $y$, and $z$ directions. (a): schematic of our three-dimensional model. (b): Cartesian surface cuts of our model. Each Cartesian cut corresponds to a two-dimensional Hofstadter non-Abelian model in the symmetric gauge.}
\label{fig1}
\end{figure}

We propose a 3D non-Abelian Hofstadter model inspired by Eqs.~\eqref{eq:abelian_3d} and \eqref{eq:goldman_3d}. Specifically, we insert the three Pauli matrices from Eq.~\eqref{eq:goldman_3d} into the complex hopping phases in Eq.~\eqref{eq:abelian_3d} along the three directions.
This replacement leads to inhomogeneous, Hofstadter--Harper-like, non-Abelian gauge potentials in 3D
\begin{align} \label{eq:3d_gauge}
     \mathbf{A}_{\pm} &= \left[ \left(l \pm n \right) \theta_{x} \sigma_x, \left( m \pm l \right) \theta_{y} \sigma_y, \left(n \pm m \right) \theta_{z} \sigma_z \right].
\end{align}
The associated Hamiltonian (see Fig.~\ref{fig1}) is given by
\begin{align} \label{eq:hh_ham}
    \begin{split}
        H_{\pm}(\theta_{x},\theta_{y},\theta_{z}) = - \sum_{m,n,l} & t_x c^\dagger_{m+1,n,l} e^{\iu \left(l \pm n \right) \theta_{x} \sigma_x} c_{m,n,l} \\
        + & t_y c^\dagger_{m, n+1, l} e^{\iu \left( m \pm l \right) \theta_{y} \sigma_y} c_{m,n,l} \\
        + & t_z c^\dagger_{m, n, l+1} e^{\iu \left(n \pm m \right) \theta_{z} \sigma_z} c_{m,n,l} + \mathrm{H.c..} 
    \end{split}
\end{align}
Throughout this work, we assume that $t_x=t_y=t_z=1$.
Similar to Eq.~\eqref{eq:abelian_3d}, the choice of $\pm$ corresponds to different magnetic fields and thus different eigenstates (see Supplementary Materials Fig.~\ref{sm_fig:compfig}a and c). A crucial feature of $H_{\pm}$ is that every one of its 2D Cartesian surfaces (along the $xy$, $zx$, and $yz$ planes) reduce to a 2D non-Abelian Hofstadter model~\cite{yang2020non} (see Fig.~\ref{fig1}b). Meanwhile, all layers, along all three directions, are stacked and connected with complex non-Abelian hoppings. As a result, the system contains spin-orbit coupling among a trio of Hofstadter butterfly pairs.

Analogous to the original Hofstadter model, this three-dimensional non-Abelian system can be solved in an enlarged magnetic unit cell if $\theta_{x}$, $\theta_{y}$, and $\theta_{z}$ are rational multiples of $2 \pi$, \ie $\theta_{x} = 2 \pi p_{x} / q_{x}$, $\theta_{y} = 2 \pi p_{y} / q_{y}$, and $\theta_{z} = 2 \pi p_{z} / q_{z}$ (where $p_i$ and $q_i$ are co-prime). The size of the magnetic unit cell is $\mathrm{lcm} \left( q_{y}, q_{z} \right) \times \mathrm{lcm} \left( q_{x}, q_{z} \right) \times \mathrm{lcm} \left( q_{x}, q_{y} \right) \equiv Q_{x} \times Q_{y} \times Q_{z}$, where $\mathrm{lcm}$ denotes the least common multiple. Consequently, the associated magnetic Brillouin zone is $k_x \in \left[ 0, 2 \pi / Q_{x} \right)$, $k_y \in \left[ 0, 2 \pi / Q_{y} \right)$, $k_z \in \left[ 0, 2 \pi / Q_{z} \right)$.

Although $H_\pm$ is non-Abelian in general, there are situations (\eg the obvious case $\theta_{x} = \theta_{y} = \theta_{z} = 0$) when the model reduces to Abelian. Therefore, to obtain the necessary and sufficient condition under which the models are genuinely non-Abelian~\cite{goldman2014light,yang2020non}, we examine the commutativity of unit plaquette loop operators at arbitrary lattice sites in three directions
\begin{align} \label{2.5} 
    \bm{W}^{\mu\nu}_\mathbf{r} = U_\nu^\dagger \left( \mathbf{r} \right) U_\mu^\dagger \left( \mathbf{r} + \hat e_\nu \right) U_\nu \left( \mathbf{r} + \hat e_\mu \right) U_\mu \left( \mathbf{r} \right),
\end{align}
where $\left\{\mu,\nu\right\}=\left\{x,y,z\right\}$, $\hat{e}_\mu$ is the unit vector in the $\mu$ direction, and we adopt the counterclockwise convention.

Specifically, the loop operators for a unit plaquette at site $\mathbf{r}=\left( m, n, l \right)$ are given by
\begin{subequations} \label{eq:loops}
    \begin{align}
        \bm{W}^{xy,\pm}_\mathbf{r}  &= \Theta_{y}^{- \left(m \pm l \right)} \Theta_{x}^{- \left[ l \pm \left( n + 1 \right) \right]} \Theta_{y}^{m+1 \pm l} \Theta_{x}^{l \pm n} \label{2.10} \\
        \bm{W}^{zx,\pm}_\mathbf{r}  &= \Theta_{x}^{- \left( l \pm n \right)} \Theta_{z}^{- \left[ n \pm \left( m + 1 \right) \right]} \Theta_{x}^{l+1 \pm n} \Theta_{z}^{n \pm m} \label{2.11} \\
        \bm{W}^{yz,\pm}_\mathbf{r}  &= \Theta_{z}^{- \left( n \pm m \right)} \Theta_{y}^{- \left[ m \pm \left( l+1 \right) \right]} \Theta_{z}^{n+1 \pm m} \Theta_{y}^{m \pm l} \label{2.12}.
    \end{align}
\end{subequations}
where $\Theta_{x}^m \equiv \exp \left( \iu m \theta_{x} \sigma_x \right)$, $\Theta_{y}^m \equiv \exp \left( \iu m \theta_{y} \sigma_y \right)$, and $\Theta_{z}^m \equiv \exp \left( \iu m \theta_{z} \sigma_z \right)$. We also define $\Theta_{x} \equiv \Theta_{x}^1$, $\Theta_{y} \equiv \Theta_{y}^1$, and $\Theta_{z} \equiv \Theta_{z}^1$ for compact notation. $\pm$ in the superscript denotes the choice of gauge fields in Eq.~\eqref{eq:3d_gauge}. We prove below that both $H^+$ and $H^{-}$ reduce to Abelian, \ie Eqs.~\eqref{eq:loops} are Abelian, if and only if 
\begin{align} \label{eq:condition}
    \text{at least two of } \theta_{x}, \theta_{y}, \text{ and } \theta_{z} \text{ are either } 0 \text{ or } \pi.
\end{align}
We consider loop operators near $\mathbf{R}=\mathbf{0}$ for $H_{-}$, which are given by
\begin{align} \label{eq:plaquettes_near_zero}
    \begin{aligned}
        {W}^{xy,-}_{0,0,0} = \Theta_{x}^{\mp 1} \Theta_{y}, 
        {W}^{xy,-}_{0,0,\mp 1} = \Theta_{y} \Theta_{x}^{\mp 1}, 
        {W}^{zx,-}_{0,0,0} = \Theta_{z}^{\mp 1} \Theta_{x}, \\
        {W}^{zx,-}_{0,\mp 1,0} = \Theta_{x} \Theta_{z}^{\mp 1}, 
        {W}^{yz,-}_{0,0,0} = \Theta_{y}^{\mp 1} \Theta_{z}, 
        {W}^{yz,-}_{\mp 1,0,0} = \Theta_{z} \Theta_{y}^{\mp 1}.
    \end{aligned}
\end{align}
To derive a necessary condition, we note that all of these loop operators must commute. Since these loop operators deal with permutations of $\theta_{x}$, $\theta_{y}$, and $\theta_{z}$, we adopt  $\left\{a,b,c\right\} \in \left\{x,y,z\right\}$ to denote their permutations. 
Eq.~\eqref{eq:plaquettes_near_zero} requires
\begin{align} \label{sm_eq:commutation}
    \left[\Theta_a\Theta_b, \Theta_c\Theta_a \right] = 0.
\end{align}
We can evaluate this commutator explicitly as
\begin{widetext}
    \begin{align}
        \begin{split}
            -2 \epsilon_{abc} \cos \left( 2 \theta_a \right) \sin \left( \theta_b \right) \sin \left( \theta_c \right) \sigma_a \\
            + \left( 2 \sin^2 \left( \theta_a \right) \sin \left( \theta_b \right) \cos \left( \theta_c \right) + 2 \epsilon_{abc} \sin \left( \theta_a \right) \cos \left( \theta_a \right) \cos \left( \theta_b \right) \sin \left( \theta_c \right) \right) \sigma_b \\
            + \left( 2 \sin^2 \left( \theta_a \right) \cos \left( \theta_b \right) \sin \left( \theta_c \right) + 2 \epsilon_{abc} \sin \left( \theta_a \right) \cos \left( \theta_a \right) \sin \left( \theta_b \right) \cos \left( \theta_c \right) \right) \sigma_c
        \end{split}
         = 0,
    \end{align}
\end{widetext}
where $\epsilon_{abc}$ is the Levi-Civita symbol.
The coefficients for all Pauli matrices must vanish to satisfy Eq.~\eqref{sm_eq:commutation}. Taken together, we must have
\begin{subequations}
    \begin{align}
        \cos \left( 2 \theta_a \right) \sin \left( \theta_b \right) \sin \left( \theta_c \right) &= 0 \label{sm_eq:cond1} \\
        \sin^2 \left( \theta_a \right) \sin \left( \theta_b \right) \cos \left( \theta_c \right) &= 0 \label{sm_eq:cond2}\\
        \sin^2 \left( \theta_a \right) \cos \left( \theta_b \right) \sin \left( \theta_c \right) &= 0.\label{sm_eq:cond3}
    \end{align}
\end{subequations}
Two situations arise: $\sin \left( \theta_a \right) = 0$ or $\sin \left( \theta_a \right) \neq 0$. 

If $\sin \left( \theta_a \right) = 0$, $\cos \left( 2 \theta_a \right) \neq 0$, at least one of $\sin \left( \theta_b \right)$ and $\sin \left( \theta_c \right)$ is 0 by Eq.~\eqref{sm_eq:cond1}. Thus, at least two of $\theta_a$, $\theta_b$, and $\theta_c$ are either 0 or $\pi$. This is a necessary condition.

If $\sin \left( \theta_a \right) \neq 0$, Eqs.~\eqref{sm_eq:cond2} and \eqref{sm_eq:cond3} imply that either $\sin \left( \theta_b \right) = \sin \left( \theta_c \right) = 0$ or $\cos \left( \theta_b \right) = \cos \left( \theta_c \right) = 0$. The former condition is equivalent to the necessary condition above. For the latter, $\sin \left( \theta_b \right) \sin \left( \theta_c \right) \neq 0$ requires $\cos \left( 2 \theta_a \right)= 0$ [as per Eq.~\eqref{sm_eq:cond1}], \ie $\cos \left( 2 \theta_{x} \right) = \cos \left( 2 \theta_{y} \right) = \cos \left( 2 \theta_{z} \right) = 0$. As a result, we cannot satisfy $\cos \left( \theta_b \right) = \cos \left( \theta_c \right) = 0$ anymore. Thus, the latter case results in a contradiction.

So far, we have proven that our condition Eq.~\eqref{eq:condition} is necessary for $H_{-}$ being Abelian. We now examine the $H_+$. In this case, the operators are in the form $\Theta_a^{-1} \Theta_b$ or $\Theta_b\Theta_a^{-1}$. The associated loop operators must commute, \ie $\left[ \Theta_a^{-1} \Theta_b, \Theta_c^{-1} \Theta_a \right] = 0$ for all choices of distinct $a$, $b$, and $c$. We again evaluate the commutator explicitly:
\begin{widetext}
    \begin{align}
        \begin{split}
            2 \epsilon_{abc} \sin \left( \theta_b \right) \sin \left( \theta_c \right) \sigma_a \\
            + \left( -2 \sin^2 \left( \theta_a \right) \sin \left( \theta_b \right) \cos \left( \theta_c \right) + 2 \epsilon_{abc} \sin \left( \theta_a \right) \cos \left( \theta_a \right) \cos \left( \theta_b \right) \sin \left( \theta_c \right) \right) \sigma_b \\
            + \left( 2 \sin^2 \left( \theta_a \right) \cos \left( \theta_b \right) \sin \left( \theta_c \right) + 2 \epsilon_{abc} \sin \left( \theta_a \right) \cos \left( \theta_a \right) \sin \left( \theta_b \right) \cos \left( \theta_c \right) \right) \sigma_c
        \end{split}
         = 0.
    \end{align}
\end{widetext}
We can follow the same arguments as those of $H_{-}$ to arrive at the same necessary condition.

Therefore, we have proven that at least two of $\theta_{x}$, $\theta_{y}$, and $\theta_{z}$ being either 0 or $\pi$ is a necessary Abelian condition. To prove it is also a sufficient condition, let $\theta_a$ and $\theta_b$ be 0 or $\pi$. $\Theta^n_a$ and $\Theta_b^n$ consequently reduce to $\pm 1$ for any integer $n$, which always commute with arbitrary SU(2) phase factors. The remaining link variables, $\Theta_c^n$, is an Abelian group as they are exponentials of a single Pauli matrix. Taken together, the necessary condition is also sufficient, rendering it the genuine Abelian condition of $H_\pm$. Recalling that the genuine non-Abelian condition of the associated 2D model~\cite{yang2020non} requires non-divisibility of the gauge potentials by $\pi$, the 3D model studied here therefore becomes genuinely Abelian if at least one of its Cartesian surfaces (see Fig.~\ref{fig1}b) is non-Abelian.

\begin{figure}[htbp]
    \includegraphics[width=\columnwidth]{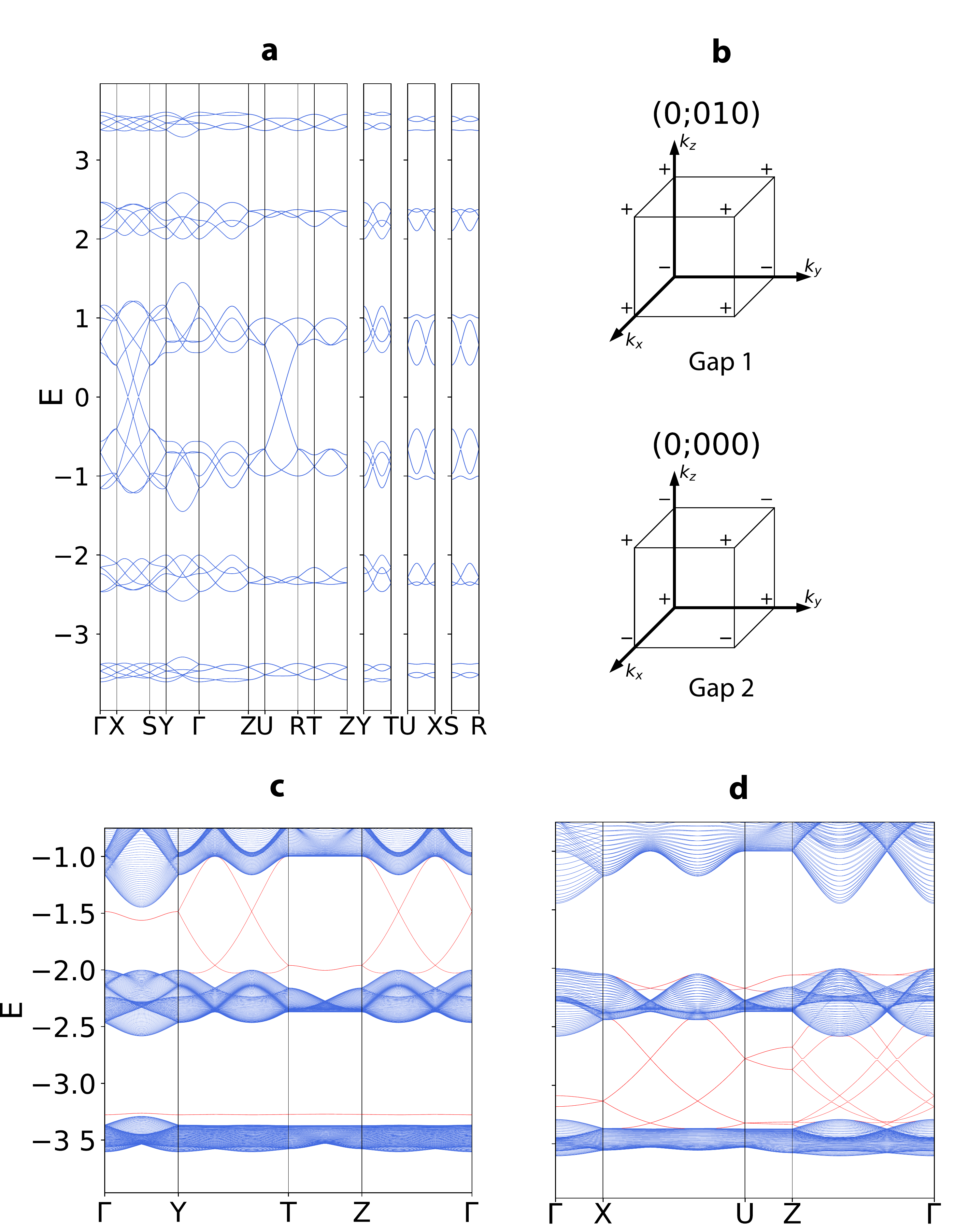}
    \caption{\textbf{a.} bulk band structure of our model with $\left( \theta_{x}, \theta_{y}, \theta_{z} \right) = \left( 0, \pi, 2 \pi / 3 \right)$ sampled along paths between high-symmetry points in the MBZ. \textbf{b.} Inversion eigenvalues at TRIMs. \textbf{c-d.} Surface band structure of the model cut along the $x$ (c) and $y$ (d) directions (with 20 magnetic unit cells). Surface states are highlighted in red.}
    \label{fig:weak}
\end{figure}

\begin{table}[ht]
    \caption{\label{tab:table2}%
    The strong and weak $\mathbb{Z}_2$ indices for $\left( \theta_{x}, \theta_{y}, \theta_{z} \right) = \left( 0, \pi, 2 \pi / 3 \right)$ for the lowest two gaps are shown. The $\mathbb{Z}_2$ indices for the two higher gaps follow by chiral symmetry.}
    \begin{ruledtabular}
        \begin{tabular}{llllll}
            Gap & Bands     & $\nu_0$ & $\nu_1$ & $\nu_2$ & $\nu_3$ \\
            \colrule
            1   & 12        & 0       & 0       & 0       & 0       \\
            2   & 24        & 0       & 0       & 1       & 0
        \end{tabular}
    \end{ruledtabular}
    \label{tab:weak}
\end{table}

\begin{table}[ht]
    \caption{\label{tab:strong}%
    The strong and weak $\mathbb{Z}_2$ indices for $\left( \theta_{x}, \theta_{y}, \theta_{z} \right) = \left( 2 \pi / 3, 2 \pi / 3, 0 \right)$ for each gap is shown. Although this choice of gauge fields does not enable chiral symmetry in the system, the sublattice symmetry remains intact, which connects the topology of positive and negative bandgaps (\eg see inversion eigenvalues of Gap 3 and Gap 6 in Fig.~\ref{fig:strong}c).}
    \begin{ruledtabular}
        \begin{tabular}{llllll}
            Gap          & Bands   & $\nu_0$ & $\nu_1$ & $\nu_2$ & $\nu_3$ \\
            \colrule
            1            & 6       & 1       & 1       & 1       & 1       \\
            2            & 12      & 1       & 1       & 1       & 1       \\
            3 (complete) & 18      & 1       & 1       & 1       & 1       \\
            4            & 24      & 0       & 1       & 1       & 1       \\
            5            & 30      & 0       & 1       & 1       & 1       \\
            6 (complete) & 36      & 1       & 0       & 0       & 0       \\
            7            & 42      & 1       & 0       & 0       & 0       \\
            8            & 48      & 1       & 0       & 0       & 0
        \end{tabular}
    \end{ruledtabular}
\end{table}

$H_\pm$ obeys time-reversal symmetry $i \sigma_y K$ and inversion symmetry $P$. Therefore, its spectrum consists of Kramers doublets in the entire magnetic Brillouin zone. As the cubic lattices are bipartite, the model obeys a sublattice symmetry that maps $E \left( k_x, k_y, k_z \right) \rightarrow -E \left( k_x + \pi, k_y + \pi, k_z + \pi \right)$. When at least one of $q_{x}$, $q_{y}$, and $q_{z}$ is even, \ie $q_{x}q_{y}q_{z}$ is even, $H_{\pm}$ also respects chiral symmetry, which we prove by leveraging the Harper's equation of $H_\pm$ in Supplementary Materials~\sisec{sm_sec:chiral}. We obtain the explicit form of the chiral operator as follows. Without loss of generality, we assume $Q_{x} = \mathrm{lcm} \left( q_{y}, q_{z} \right)$ is even, The chiral operator $S_{x}$ is
\begin{align} \label{2.161}
    \left( S_{x} u \right)_{m,n,l} = \left( -1 \right)^{m + n \alpha_{xy} + l \alpha_{xz}} \left( \iu \right)^{Q_{x} / 2} \sigma_0 u_{m + Q_{x} / 2, n, l},
\end{align}
where $u$ is the wavefunction. Here, we define $\alpha_{\mu\nu} \equiv \left( Q_\mu p_\nu / q_\nu + 1 \right) \mod 2$. 
$S_{y}$ and $S_{z}$ can be defined similarly if $Q_{y}$ and $Q_{z}$ are respectively even. The form of this 3D chiral operator is reminiscent of those in the 2D Abelian~\cite{wen1989winding} and non-Abelian~\cite{yang2020non} Hofstadter models.

\begin{figure}[htbp!]
    \includegraphics[width=\columnwidth]{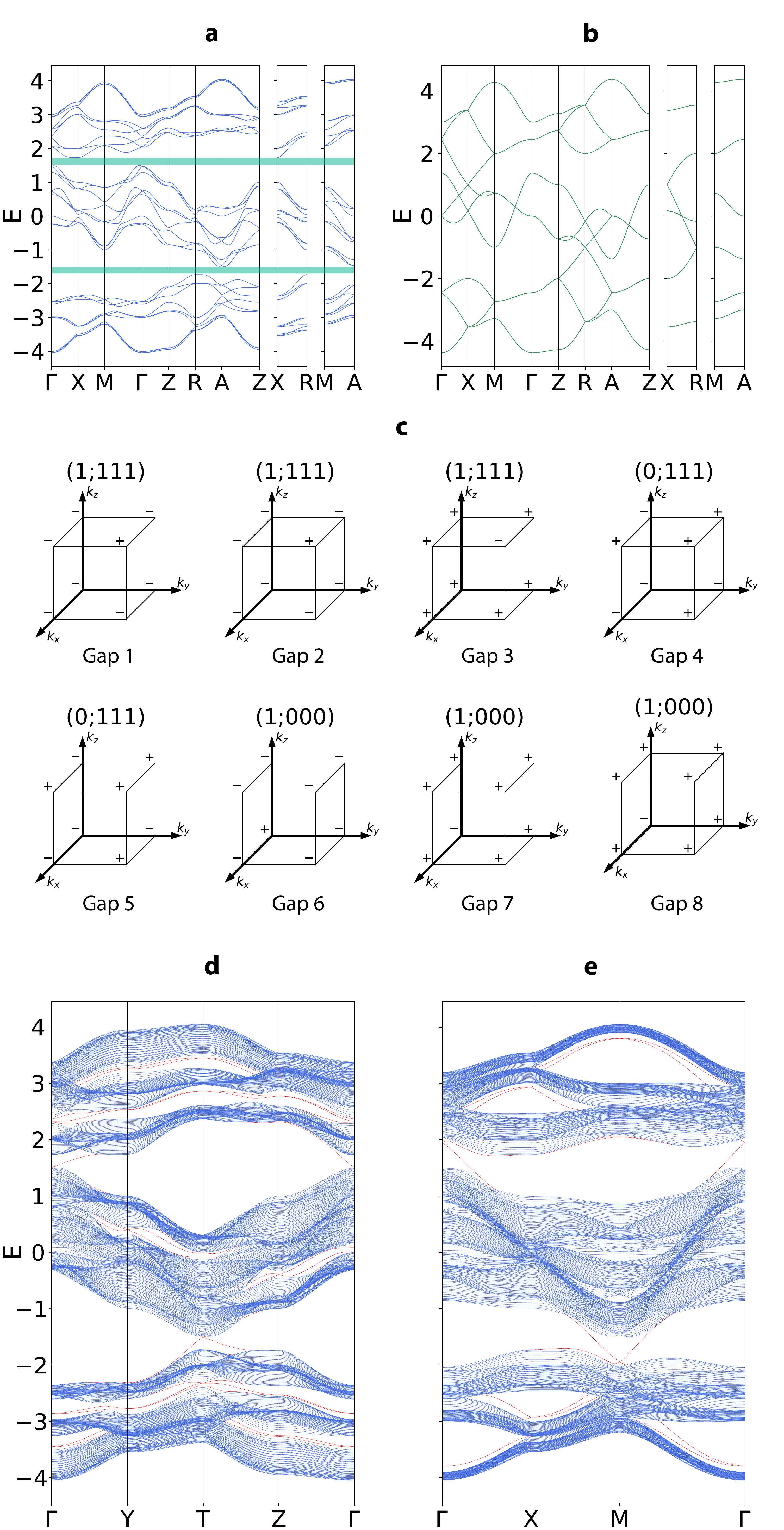}
    \caption{\textbf{a-b.} bulk band structure of the non-Abelian (a) and Abelian (b) models with the choice of gauge fields $\left( \theta_{x}, \theta_{y}, \theta_{z} \right) = \left( 2 \pi / 3, 2 \pi / 3, 0 \right)$, sampled along paths between high-symmetry points in the MBZ: spin-orbit coupling opens band gaps (shaded green in a) that are absent in the Abelian model. \textbf{c.} Inversion eigenvalues and the $\mathbb{Z}_2$ indices for all band gaps. \textbf{d-e.} Surface band structure cut along the $x$ (d) and $z$ (e) directions (with 20 magnetic unit cells) with remaining directions periodic. Surface states are highlighted in red.}
    \label{fig:strong}
\end{figure}

Finding the eigenspectrum of the model is computationally expensive as the system size increases rapidly with $q_{x}$, $q_{y}$, and $q_{z}$---the size of the magnetic unit cell is $\mathrm{lcm} \left( q_{y}, q_{z} \right) \times \mathrm{lcm} \left( q_{z}, q_{x} \right) \times \mathrm{lcm} \left( q_{x}, q_{y} \right)$. In the following, we focus on $H_+$ and drop the subscript hereafter. We study two choices of gauge fields, $\left(\theta_x,\theta_y,\theta_z\right)=\left( 2 \pi / 3, 2 \pi / 3, 0 \right)$ and $\left(\theta_x,\theta_y,\theta_z\right)=\left( 2 \pi / 3, 2 \pi / 3, 0 \right)$, which lie within the Abelian and non-Abelian regimes, respectively, according to our genuine condition proven above. For $\left( \theta_{x}, \theta_{y}, \theta_{z} \right) = \left( 0, \pi, 2 \pi / 3 \right)$, the magnetic unit cell has a dimension of $6 \times 3 \times 2$ for a total of 72 bands, \ie 36 Kramers partners. This choice of gauge fields lies within the Abelian regime: therefore, the eigenspectrum of $H$ doubles that of $H_1$. Regarding gapped phases, there are a total of four bandgaps, half of which are chiral partners of the other half. As a result, we only need to examine the lowest two bandgaps, whose strong and weak $\mathbb{Z}_2$ indices $\left( \nu_0 ; \nu_1, \nu_2, \nu_3 \right)$~\cite{fu2007topological} (see Supplementary materials) are shown in Table \ref{tab:table2}, obtained via inversion eigenvalues at the eight time-reversal invariant momenta (see Fig.~\ref{fig:weak}b). Because all its $\mathbb{Z}_2$ indices simultaneously vanish, the first gap around $E=-3$ is topologically trivial, as also supported by its trivial surface states in Fig.~\ref{fig:weak}. On the other hand, the second gap, around $E=-1.5$, is a weak topological insulator (TI) (see Table~\ref{tab:weak}) with an odd index $\nu_2=1$. We confirm this weak-TI diagnosis with surface $x$-cut (Fig.~\ref{fig:weak}c) and $y$-cut (Fig.~\ref{fig:weak}d) calculations. The surface states of the second gap are non-trivial and trivial in the $x$-cut and $y$-cut systems, respectively. Meanwhile, the surface states of the first gap are trivial in both truncation directions. These surface states are consistent with the bulk diagnosis in Table~\ref{tab:weak}.

Next, we study $H_+\left( 2 \pi / 3, 2 \pi / 3, 0 \right)$. Evidently, this choice of hopping phases lies within the genuine non-Abelian regime. To highlight the associated consequence, we compare the bulk spectra of $H_+$ with that of the three-dimensional Abelian Hofstadter model $H_1\left( 2 \pi / 3, 2 \pi / 3, 0 \right)$ [Eq.~\eqref{eq:abelian_3d}], as shown in Fig.~\ref{fig:strong}a and b, respectively, sampled along high-symmetry lines in the three-dimensional magnetic Brillouin zone. Compared to $H_1$ in Fig.~\ref{fig:strong}b, which is gapless, new band gaps (with complete band gaps shown in shaded blue) are opened in Fig.~\ref{fig:strong}a due to the addition of spin-orbit coupling. Similar band-gap opening also appears for other choices of gauge fields (see Supplementary Fig.~\ref{sm_fig:bandgap}).
For $\left( \theta_{x}, \theta_{y}, \theta_{z} \right) = \left( 2 \pi / 3, 2 \pi / 3, 0 \right)$, the magnetic unit cell has dimensions $3 \times 3 \times 3$ with a total of 27 Kramers pairs. There are a total of eight band gaps with two complete ones (highlighted by green shadings in Fig.~\ref{fig:strong}). The strong and weak $\mathbb{Z}_2$ indices $\left( \nu_0 ; \nu_1, \nu_2, \nu_3 \right)$ are also evaluated for all band gaps using inversion eigenvalues (Fig.~\ref{fig:strong}c) and shown in Table~\ref{tab:strong}. Evidently, the system is a strong 3D topological insulator at both the complete band gaps (namely, Gap 3 and Gap 6). We also include the calculation of the Wannier spectrum with odd winding~\cite{soluyanov2011} (see Supplementary Materials Fig.~\ref{fig:wannierfig}), which affirms our results for $\mathbb{Z}_2$ invariants calculated using inversion eigenvalues. We verify such bulk analysis by calculating the surface spectra with open boundary conditions in the $x$ (Fig.~\ref{fig:strong}d) and $z$ ((Fig.~\ref{fig:strong}e) directions. In contrast to those of the weak insulating phase shown in Fig.~\ref{fig:weak}, the complete gaps of the strong insulating phase exhibit helical surface states under both types of truncation. Notably, with $x$-cut (Fig.~\ref{fig:strong}d), the surface Dirac points appear at T and $\Gamma$ points for Gap 3 and Gap 6, respectively, which is ensured by the sublattice symmetry of the Hamiltonian. A similar correspondence between the surface Dirac points appears also for the $z$-cut spectrum in Fig.~\ref{fig:strong}e.

In conclusion, we have introduced a three-dimensional non-Abelian generalization of the Hofstadter model with three spatially inhomogeneous and linearly varying gauge fields on a cubic lattice, proven the genuine non-Abelian condition of the model, analyzed its internal symmetries, and discussed the strong and weak $\mathbb{Z}_2$ insulating phases under different choices of gauge fields.  Experimentally, it may be possible to realize the models on various platforms, including photonic coupled waveguide/resonator arrays and synthetic frequency combs, topological circuit systems, and spin-orbit-coupled atomic gases. Future directions also include analyzing the rich crystalline symmetries of the model and identifying the associated first-order and higher-order crystalline phases.

\appendix
\renewcommand{\thesubsection}{\Alph{section}.\Roman{subsection}}
\renewcommand{\thesubsection}{\Roman{subsection}}
\renewcommand{\thefigure}{A\arabic{figure}}
\setcounter{section}{0}
\setcounter{subsection}{0}
\setcounter{figure}{0}

We thank Liang Fu, Hoi Chun Po, and Ashvin Vishwanath for discussions. This material is based upon work supported in part by the Air Force Office of Scientific Research under the award number FA9550-20-1-0115, as well as in part by the US Office of Naval Research (ONR) Multidisciplinary University Research Initiative (MURI) grant N00014-20-1-2325 on Robust Photonic Materials with High-Order Topological Protection. This material is also based upon work supported in part by the U. S. Army Research Office through the Institute for Soldier Nanotechnologies at MIT, under Collaborative Agreement Number W911NF-18-2-0048.

\bibliographystyle{apsrev4-2}
\bibliography{references}

\end{document}

%-----TITLE-----
\title{
Supplementary Materials\\
Three-dimensional non-Abelian generalizations of the Hofstadter model: spin-orbit-coupled butterfly trios}
\author{Vincent Liu}
\email{vincent\_liu@berkeley.edu}
\affiliation{\mitaffil}
\affiliation{\ucbaffil}
\author{Yi Yang}
\email{yiy@mit.edu}
\affiliation{\mitaffil}
\author{John D. Joannopoulos}
\affiliation{\mitaffil}
\author{Marin~Solja\v{c}i\'{c}}
\affiliation{\mitaffil}

\maketitle

%-----CHANGE SETUP FOR PARAGRAPH INDENTS AND SKIPS-----
\setlength{\parindent}{0em}
\setlength{\parskip}{.5em}

%----- MAIN MATTER ------

%-----------------------------------------------------------

\section{Additional bulk spectra of other choices of gauge fields in the 3D Abelian and non-Abelian Hofstadter models} \label{sm_sec:plots}

\begin{figure*}[htpb!]
    \includegraphics[width=0.7\linewidth]{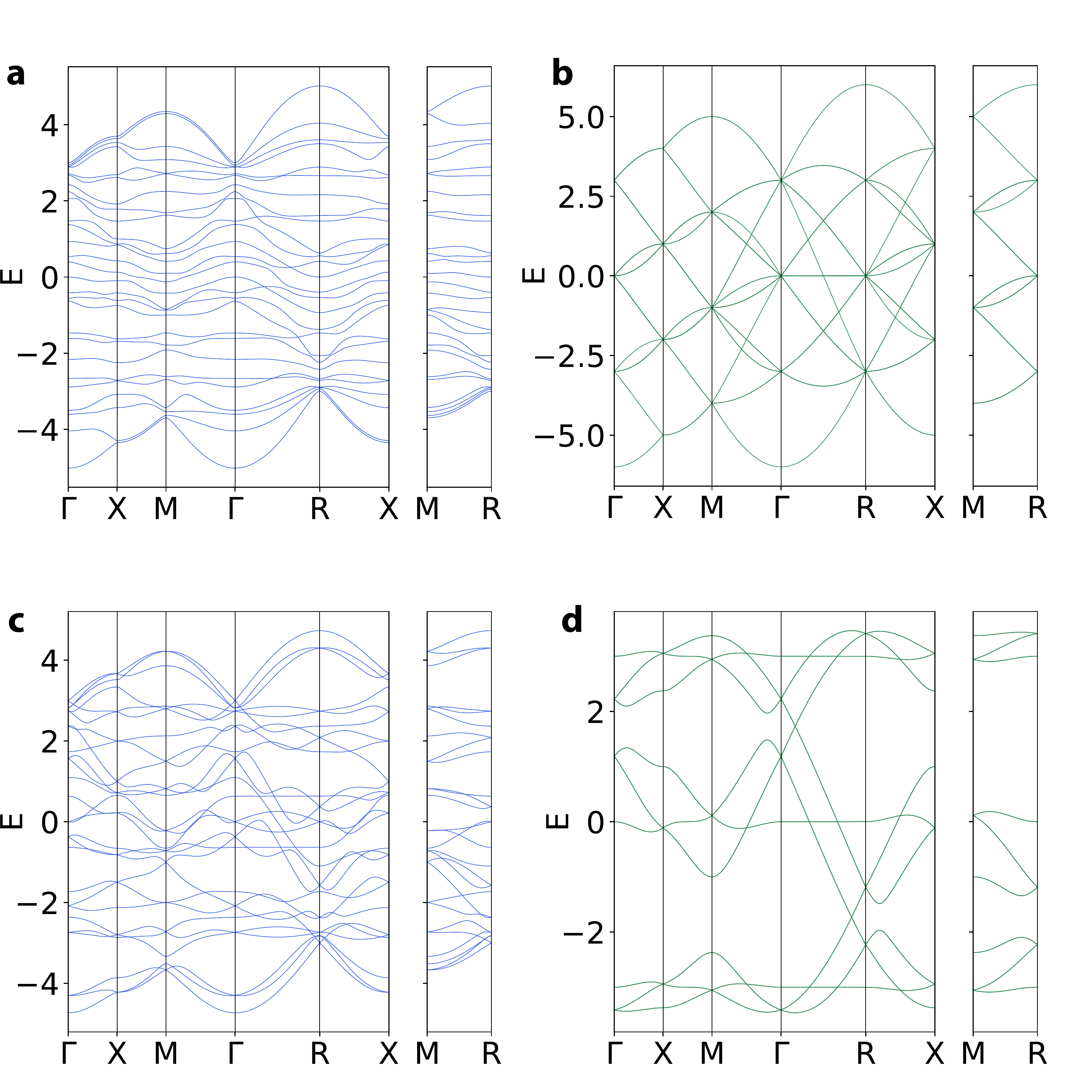}
    \caption{Bulk band structure with $\left( \theta_{x}, \theta_{y}, \theta_{z} \right) = \left( 2 \pi / 3, 2 \pi / 3, 2 \pi / 3 \right)$, sampled along paths between high-symmetry points in the MBZ. Non-Abelian models are shown on the left (blue) while Abelian models are shown on the right (green). (a) and (b) take the + choice, while (c) and (d) take the - choice in $H$ [Eq.~\eqref{eq:3d_gauge}] and $H_1$ [Eq.~\eqref{eq:abelian_3d}], respectively. It is evident that all models show distinct bulk spectra because of their different associated magnetic fields. Bandgaps only appear in a.}
    \label{sm_fig:compfig}
\end{figure*}

\begin{figure*}[htpb!]
    \includegraphics[width=0.7\linewidth]{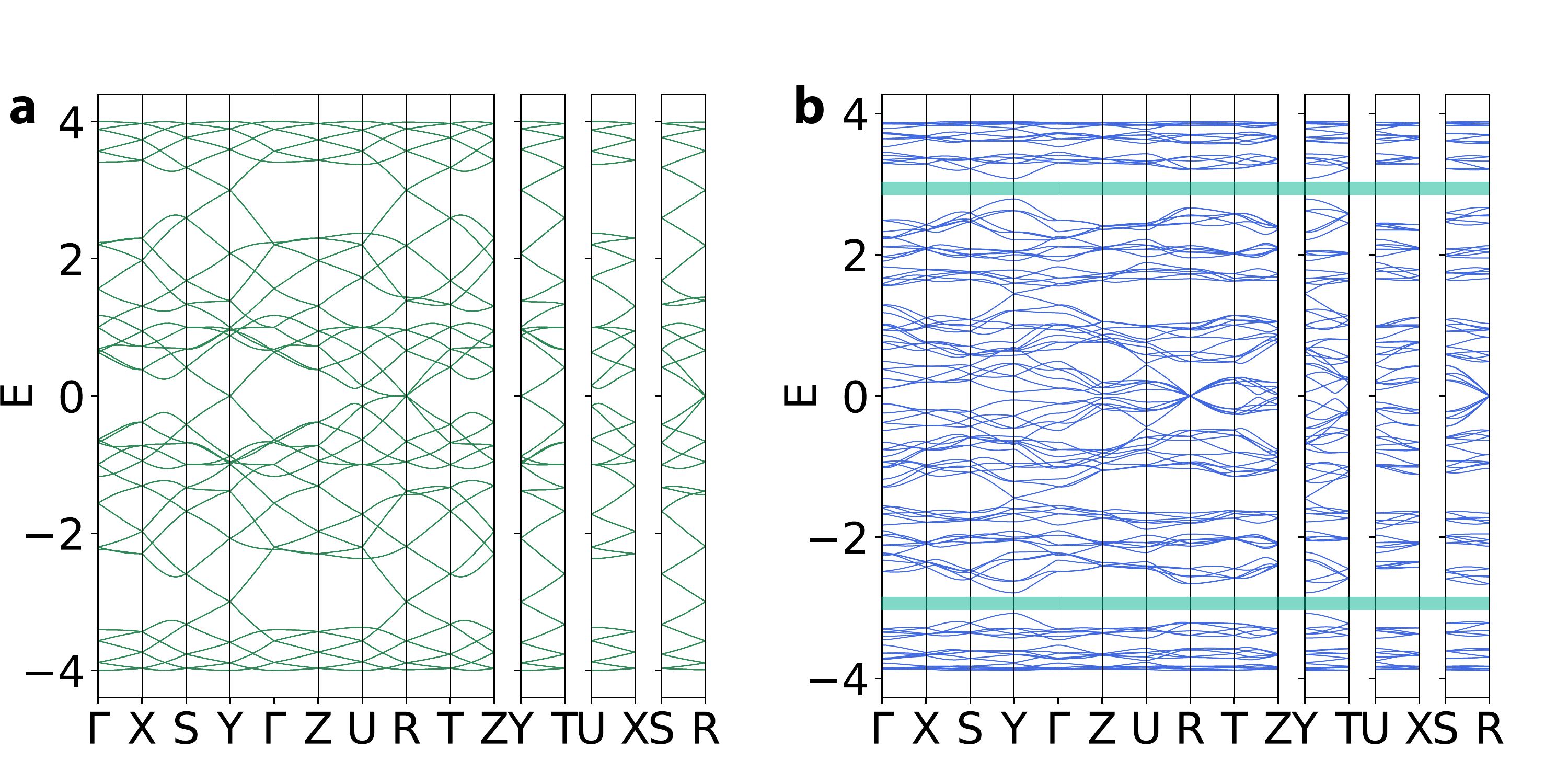}
    \caption{Bulk spectra for $\left( \theta_{x}, \theta_{y}, \theta_{z} \right) = \left( \pi, 2 \pi / 3, \pi / 3 \right)$ with (a) being Abelian and (b) being non-Abelian. Similarly to the $\left( \theta_{x}, \theta_{y}, \theta_{z} \right) = \left( 0, 2 \pi / 3, 2 \pi / 3 \right)$ case, new band gaps are opened. Both Abelian and non-Abelian models respect chiral symmetry.}
    \label{sm_fig:bandgap}
\end{figure*}

\begin{figure*}[htpb!]
    \includegraphics[width=0.4\linewidth]{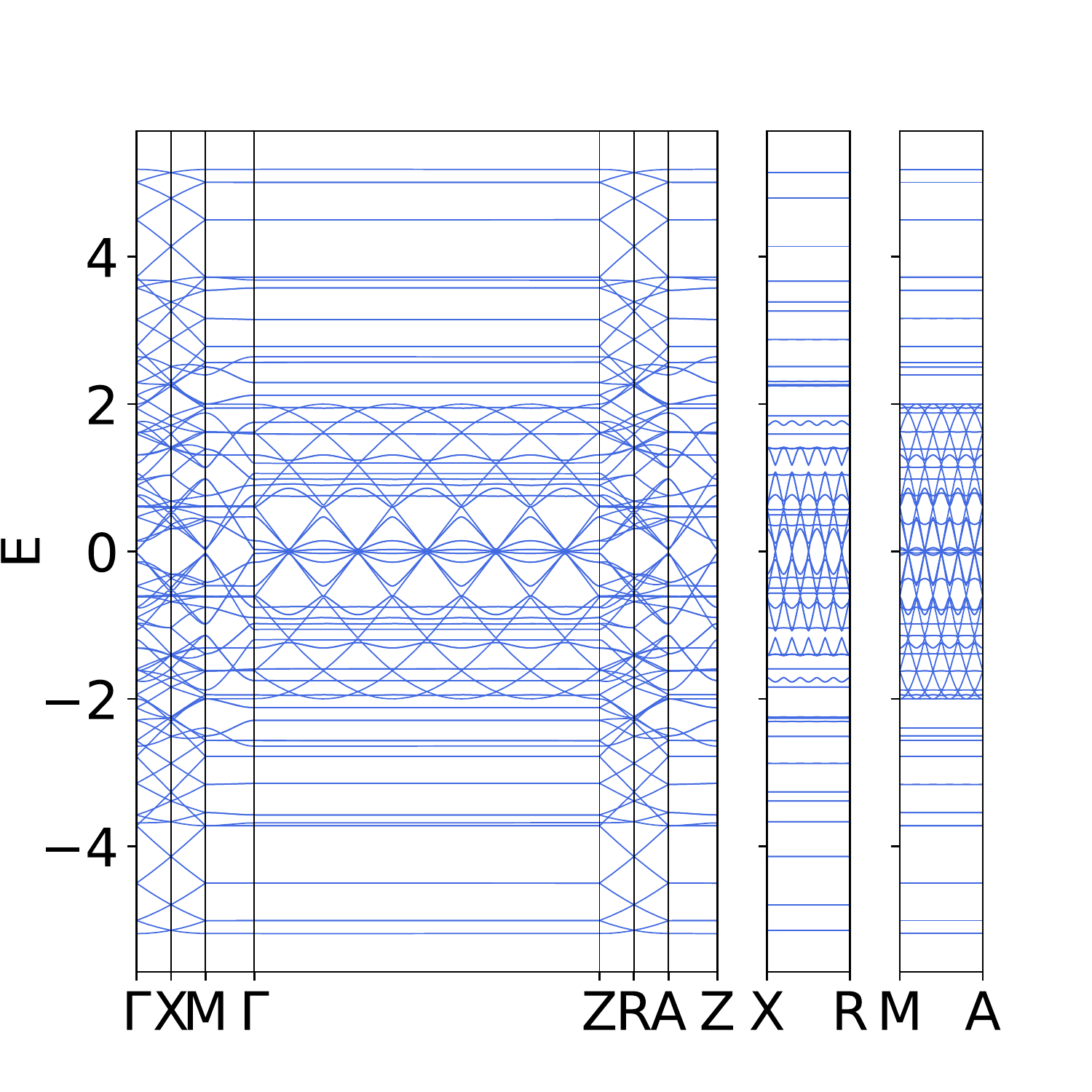}
    \caption{Bulk spectra for $\left( \theta_{x}, \theta_{y}, \theta_{z} \right) = \left( 0, 0, \pi / 5 \right)$ in the Abelian regime. Under small magnetic flux, Landau levels appear as in the original Hofstadter model.}
    \label{sm_fig:flat}
\end{figure*}

\section{Existence of Chiral Symmetry} \label{sm_sec:chiral}

In this section we prove that our model has chiral symmetry when $q_x q_y q_z$ is even. As outlined in the main text, we prove $E \left( k_x, k_y \right) \rightarrow -E \left( k_x + \pi, k_y + \pi \right)$ in general, and $E \left( k_x, k_y, k_z \right) \rightarrow -E \left( k_x + \pi, k_y + \pi, k_z + \pi \right)$ when $q_x q_y q_z$ is even.

To prove the first statement, we write the eigenvalue equation 
\begin{align} \label{b.1}
    \begin{split}
        E \Psi_{m,n,l} = &-t_x \left( \Theta^{l \pm n}_x \Psi_{m+1,n,l} + \Theta^{- \left( n \pm l \right)}_x \Psi_{m-1,n,l} \right) \\
        &- t_y \left( \Theta^{m \pm l}_y \Psi_{m,n+1,l} + \Theta^{- \left( m \pm l \right)}_y \Psi_{m,n-1,l} \right) \\
        &- t_z \left( \Theta^{n \pm m}_z \Psi_{m,n,l+1} + \Theta^{- \left( n \pm m \right)}_z \Psi_{m,n,l-1} \right).
    \end{split}
\end{align}
Using Bloch's theorem, we have $\Psi_{m,n,l} = e^{i k_x m} e^{i k_y n} e^{i k_z l} u_{m,n,l}$ for some $\left( k_x, k_y, k_z \right)$ in the magnetic Brillouin zone and $u$ periodic in the magnetic unit cell ($u_{m,n,l} = u_{m+Q_x,n,l} = u_{m,n+Q_y,l} = u_{m,n,l+Q_z}$). Making this substitution gives rise to the Harper equation
\begin{align} \label{b.2}
    \begin{split}
        E u_{m,n,l} = &-t_x \left( e^{i k_x} \Theta^{l \pm n}_x u_{m+1,n,l} \right. \\
        &\qquad \quad \left. + e^{-i k_x} \Theta^{- \left( n \pm l \right)}_x u_{m-1,n,l} \right) \\
        &- t_y \left( e^{i k_y} \Theta^{m \pm l}_y u_{m,n+1,l} \right. \\
        &\qquad \quad \left. + e^{-i k_y} \Theta^{- \left( m \pm l \right)}_y u_{m,n-1,l} \right) \\
        &- t_z \left( e^{i k_z} \Theta^{n \pm m}_z u_{m,n,l+1} \right. \\
        &\qquad \quad \left. + e^{-i k_z} \Theta^{- \left( n \pm m \right)}_z u_{m,n,l-1} \right).
    \end{split}
\end{align}
Multiplying by $e^{i \pi}$, we get
\begin{align} \label{b.3}
    \begin{split}
        -E u_{m,n,l} = &-t_x \left( e^{i \left( k_x + \pi \right)} \Theta^{l \pm n}_x u_{m+1,n,l} \right. \\
        &\qquad \quad \left. + e^{-i \left( k_x + \pi \right)} \Theta^{- \left( n \pm l \right)}_x u_{m-1,n,l} \right) \\
        &- t_y \left( e^{i \left( k_y + \pi \right)} \Theta^{m \pm l}_y u_{m,n+1,l} \right. \\
        &\qquad \quad \left. + e^{-i \left( k_y + \pi \right)} \Theta^{- \left( m \pm l \right)}_y u_{m,n-1,l} \right) \\
        &- t_z \left( e^{i \left( k_z + \pi \right)} \Theta^{n \pm m}_z u_{m,n,l+1} \right. \\
        &\qquad \quad \left. + e^{-i \left( k_z + \pi \right)} \Theta^{- \left( n \pm m \right)}_z u_{m,n,l-1} \right).
    \end{split}
\end{align}
Comparing Eq. \ref{b.2} to Eq. \ref{b.3}, we evidently have a ``translation'' symmetry $E \left( k_x, k_y, k_z \right) \rightarrow -E \left( k_x + \pi, k_y + \pi, k_z + \pi \right)$. This symmetry is the same as that of the original Hofstadter models as well as the two-dimensional models discussed in Ref.~\cite{yang2020non}.

We now show that $E \left( k_x, k_y, k_z \right) \rightarrow E \left( k_x + \pi, k_y + \pi, k_z + \pi \right)$ when $q_x q_y q_z$ is even, or when at least one of $q_x$, $q_y$, and $q_z$ is even. Suppose, without loss of generality, that $q_x$ is even. Therefore, $Q_y = \mathrm{lcm} \left( q_z, q_x \right)$ and $Q_z = \mathrm{lcm} \left( q_x, q_y \right)$ are both even. Because the magnetic Brillouin zone in the $y$ direction is $k_y \in \left[ 0, 2 \pi / Q_y \right)$, points $k_y$ and $k_y + \pi$ become equivalent. Similarly, $k_z$ and $k_z + \pi$ become equivalent as well. Taken together, we have $E \left( k_x, k_y, k_z \right) \rightarrow E \left( k_x, k_y + \pi, k_z + \pi \right)$.

For $k_x$, we argue that the $x$-direction link at lattice site $\left( m,n,l + q_x / 2\right)$ becomes equivalent to $e^{i \left( \left( l \pm n \right) \theta_x \sigma_x + \pi \right)}$. Therefore, for an eigenstate $\Psi_{m,n,l} \left( k_x, k_y, k_z \right)$ with some energy $E$, we have another eigenstate $\Psi_{m,n,l+q_x/2} \left( k_x + \pi, k_y, k_z \right)$ with the same energy $E$. This shows that we have $E \left( k_x, k_y, k_z \right) \rightarrow -E \left( k_x + \pi, k_y + \pi, k_z + \pi \right)$ and thus that our model has chiral symmetry when $q_x q_y q_z$ is even.

\section{$\mathbb{Z}_2$ indices} \label{sm_sec:z2}

We determine $\mathbb{Z}_2$ indices using the inversion eigenvalue representation of the  $\mathbb{Z}_2$ invariants by Fu and Kane~\cite{fu2007topological}. The $\mathbb{Z}_2$ invariants are determined by the quantities
\begin{align} \label{2.18}
    \delta_i = \prod_{m=1}^N \xi_{2 m} \left( \Gamma_i \right)
\end{align}
where $\Gamma_i$ is a time reversal invariant momentum, $m = 1 ... N$ indexes the occupied bands, and $\xi_{2 m} \left( \Gamma_i \right)$ is the inversion eigenvalue of the $2m$-th band at $\Gamma_i$. Each Kramers pair has the same inversion eigenvalue and is counted only once, hence the index $2m$. The different $\mathbb{Z}_2$ invariants $\nu_a$ are then given by the products
\begin{align} \label{2.19}
    \left( -1 \right)^{\nu_a} = \prod \delta_i
\end{align}
with different products for the different strong and weak $\mathbb{Z}_2$ invariants in three dimensions.

\section{Wannier Spectra}

Wannier spectra can be used to confirm our diagnoses of the $\mathbb{Z}_2$ indices of various gaps~\cite{soluyanov2011}. Below we show Wannier spectra for the first gap (bands 1-6) with gauge potentials $\left( \theta_x, \theta_y, \theta_z \right) = \left( 2 \pi / 3, 2 \pi / 3, 0 \right)$ to confirm its diagnosis using the inversion eigenvalues. 
\begin{figure*}[htpb]
    \includegraphics[width=\linewidth]{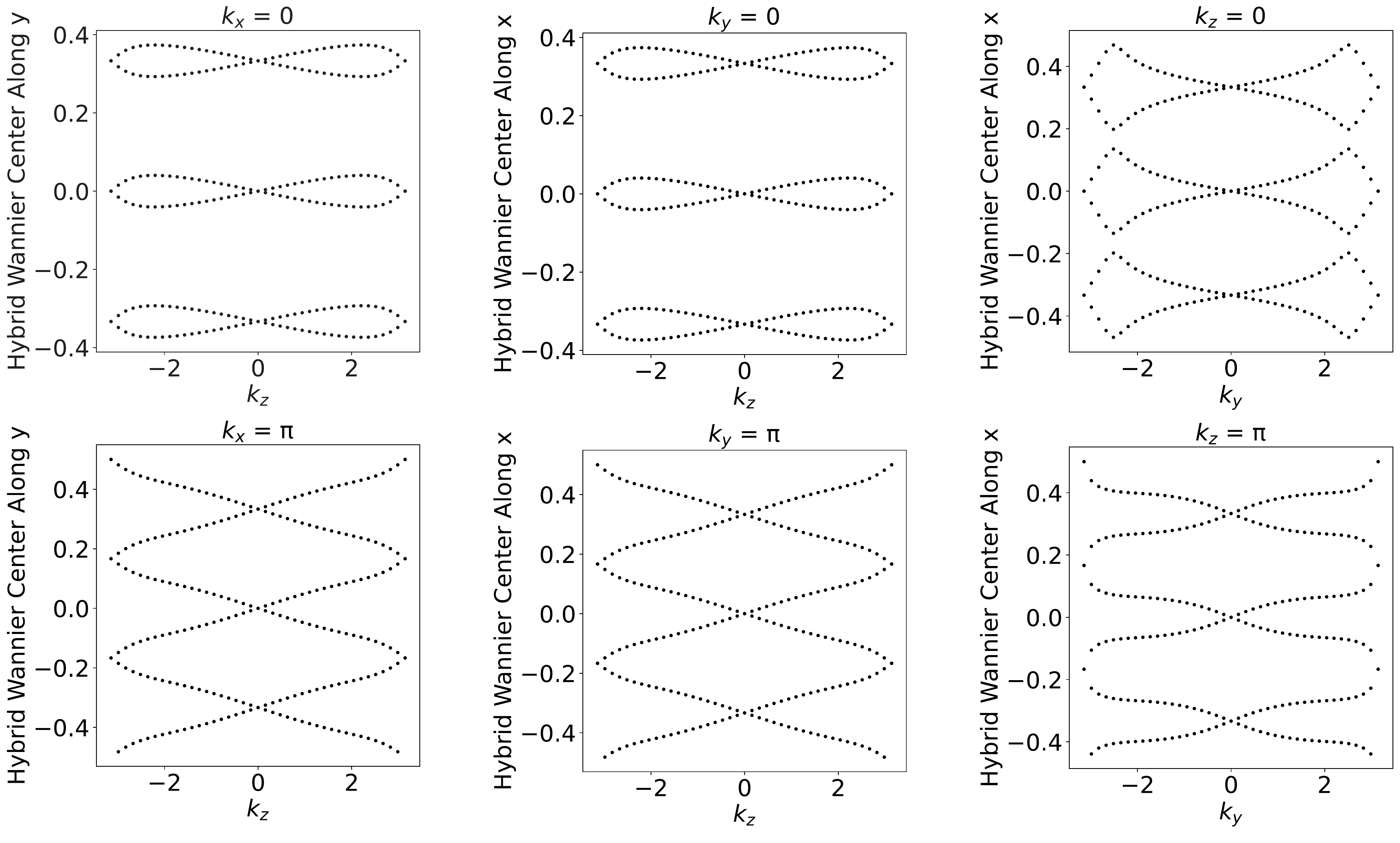}
    \caption{Wannier spectra for the first gap (bands 1-6) for our model with gauge potentials $\left( \theta_x, \theta_y, \theta_z \right) = \left( 2 \pi / 3, 2 \pi / 3, 0 \right)$. The spectra at $k_x = 0$, $k_y = 0$, and $k_z = 0$ have even times of winding, meaning $\nu_1^0 = \nu_2^0 = \nu_3^0 = 0$. In contrast, the spectra at $k_x = \pi$, $k_y = \pi$, and $k_z = \pi$ have odd times of winding, meaning $\nu_1^\pi = \nu_2^\pi = \nu_3^\pi = 1$. Thus, the $\mathbb{Z}_2$ indices are $\left( \nu_0 ; \nu_1 \nu_2 \nu_3 \right) = \left( 1; 1 1 1 \right)$, confirming the diagnosis in Table II.}
    \label{fig:wannierfig}
\end{figure*}

\newpage

\bibliographystyle{apsrev4-2}
\bibliography{references}